\documentstyle[a4,12pt]{article}

\begin{document}

\begin{center}
{\bf Quantum Mott transition in a silicon quantum dot}
\bigskip

S.\ V.\ Vyshenski\footnote{E-mail: svysh@rsfq.npi.msu.su}\dag \ddag, 
U.\ Zeitler\ddag, and 
R.\ J.\ Haug\ddag \\

\smallskip

\dag~Institute of Nuclear Physics, Moscow State University, Moscow 119899,
Russia \\ 
\ddag~Institut f\"ur Festk\"orperphysik, Universit\"at Hannover,
Appelstr. 2, D-30167 Hannover, Germany

\bigskip

February 21, 1998

\end{center}

\begin{abstract}
Considering a double-barrier structure formed by a silicon quantum dot covered
by natural oxide, we derive simple conditions for the conductance of the dot
to become a step-like function of the number of doping atoms inside 
the dot, with negligible dependence on the actual position of the
dopants. The found conditions are feasible in experimentally available
structures.  
\end{abstract}

The fabrication of $Si$ nanostructures became possible through
very recently developed new technologies \cite{Chou,Oda}.
One unique preparation technology for individual silicon quantum 
dots (SQD) has been reported in \cite{Oda}. They are spherical 
$Si$ particles with diameters $d$ in the range 5--12 nm covered
by a 1--2 nm-thick natural $SiO_2$ film. Metallic current
terminals made from degenerately doped $Si$ are defined lithographically 
to touch each individual dot from above and from below.

To ensure metallic electrodes the donor concentration $n$ 
should be $n\ge n_{Mott}$,
where $n_{Mott}=7.3\times 10^{17}$ cm$^{-3}$. The critical concentration
$n_{Mott}$ is defined by the {\em Mott criterion} \cite{Mott},
introducing the transition to a metallic type of conductivity 
in a semiconductor at:
\begin{equation}
a_B\times (n_{Mott})^{1/3}=0.27 .  \label{Mott}
\end{equation}
where $a_B$ nm is the Bohr radius of an electron  
bound to a donor inside the $Si$ crystal, in the case of
phosphorus-donors $a_B = 3$~nm \cite{Mott}.

As for the doping of the dot, the situation concerning a 
Mott transition in that small dots is much less trivial than the
one described by Eq. (\ref{Mott}).
Let us consider dots with diameters $d=10$~nm formed from n-doped
$Si$ with $n = n_{Mott}$ as an illustrative example.  
Then each dot contains in average one donor. Note that we will consider
degenerately $n^+$-doped electrodes with $n \gg n_{Mott}$ which ensures 
metallic conduction up to the borders of the dot.

Real fabrication technology \cite{Oda} provides a wafer with hundreds of
SQDs on it with current leads towards each individual SQD. Dots in average
have the same value of mean dopant concentration $n$, which is
determined by the parent material of bulk silicon the dots are formed 
from. However, on the level of each individual
SQD we will always have exactly {\em integer} number of doping atoms. If,
as in the example above, the average number of dopants 
$\overline{N_{tot}}=1$ the actual number  of donors
in the dot can have values $N_{tot} = 0, 1, 2, 3, \ldots$, with
values larger than these very unlikely.

Our objective is to illustrate, that SQDs from the same wafer fall
into several distinct sets of approximately the same conductance. The
typical value of conductance for each set is nearly completely 
determined by the number $N$ of donors present in a certain part of a SQD
so that $N$ labels each set of SQDs.

Summarizing the above, we need for a quantization of the 
conduction through a dot with $N$ donors the following conditions:

\begin{itemize}
\item Size $d$ of the dot comparable with Bohr radius: $2<d/a_B<5$.
\item Average doping $n$ of the dot $n \le d^{-3}$,
leading to a mean number of dopants  $\overline{N_{tot}} \le 1$, 
so that $N_{tot}=0, 1, 2$ are the most probable configurations 
of an individual SQD.
\item Doping of the electrodes $n_{el} \gg n_{Mott}$, 
so that current leads are perfectly metallic.
\item Dot covered by an oxide layer thick enough to suppress ballistic 
transport through the dot.
\end{itemize}

In fact all these condition can be simultaneously satisfied 
for SQD fabricated with the method mentioned above \cite{Oda}.

\section{Model system}

We use a simple model of a cubic SQD with $d > 2 a_B$ (we will use
$d=10$ nm for estimates), covered with an 
oxide layer with thickness $\delta = 2$~ nm, and contacted with current 
terminals from below and from above.
The $x$-axis is oriented from top to bottom along the current flow,
as shown in Fig. 1.

A tunneling current is injected into the dot via the oxide barrier from 
the top (source at $x=0$) and leaves the dot at the bottom 
(drain at $x=d$). Due to the presence of the oxide barriers
this current is non-ballistic and non-thermal. 
We assume that the high potential barriers associated with the 
oxide layers are not much affected by the voltage and the tunneling 
charges. We concentrate on what happens between these effective source and
drain (Fig. 2).

In the case when the dot can be regarded as an insulating system
it is reasonable to assume that the applied voltage equally
drops over the potential barriers and the dots. For simplicity
we neglect the difference of the dielectric constants of the 
oxide barriers and the dot. 
In this approximation we can introduce
an effective voltage  $V_{eff}=V (d-2 \delta) / d = 0.6 V$ 
describing the part of the total  
transport voltage $V$ applied between effective source and drain
which drops across the dot itself.

In this rude approximation we neglect the effect of spatial quantization 
upon values on the ionization energy, the
conductivity gap and material parameters of silicon.

\section{Dot without donors}

At $V_{eff}=0$ the Fermi level inside the dot 
is situated in the middle of the gap, i.e. $E_g/2$ bellow
the conduction band edge ($E_g=1.14$ eV at 300 K). 

As $V_{eff}$ grows, the bottom of the (still empty) conduction band bends
down accordingly. When the conduction band in the dot close to the drain
aligns with the Fermi level of the emitter we expect a drastic 
increase in the tunneling current. This threshold $V_{th}$ voltage (Fig. 2) 
for $V_{eff}$ is given by 
$V_{th} = E_g/(2 e)$, 
regardless of the number $N_{tot}$ of dopants in the dot
(as long as the dot is not yet metallic, of course). 
In the following we therefore limit our
studies to voltages 

\begin{equation}
\left|V_{eff} \right| \le V_{th} = E_g/(2 e) = 0.57 \mbox{ V}. 
\label{threshold}
\end{equation}

In this voltage range we have a 
$d$-thick barrier (formed by the dot) with always finite hight 
between effective source and drain. The intrinsic
concentration of electrons and holes at 300 K is $1.4 \times 10^{10}$
cm$^{-3}$. Even at this high temperature the  probability to have at 
least one intrinsic electron in a dot
with size $d=10$ nm is only $1.4 \times 10^{-8}$. 
So we would expect virtually no
current in this mode. This 
is confirmed by direct electrical tests \cite{Oda} 
of SQD with the required properties. 

\section{Single-donor channel}

Let us now consider one single 
single donor in the dot located at $x$ with ionization energy 
$E_d=0.045$ eV (for P as a donor).

The evident channel for current flow is single-electron tunneling from the
source to the empty impurity, and then from populated impurity
to the drain. This channel opens as soon as $V_{eff}$ reaches a
threshold $V_1$ leading to a step-like increase in the total
conductance of the dot. If the impurity is located near the drain, i.e.\ 
$d-a_B < x < d$ (as donor 1 in Fig. 2), then $V_1$ is given by
\begin{equation}
V_1=E_g/(2 e)-E_d= 0.525 \mbox{ V}.
\label{V1}
\end{equation}

In contrast, for an impurity located at distances $\Delta x>2 d E_d/E_g $ 
from the drain (i.e. further away than the threshold 
case of donor 2 in Fig. 2),
no additional current channel via a single impurity can be 
opened at low enough voltages defined in (\ref{threshold})
where virtually no background current is present. 
In the present case this value $\Delta x =
0.8$~nm, which returns us to the above criterion: only impurities located in
the immediate vicinity (defined within the accuracy $a_B$) of the drain
contribute to the single-impurity channel. 

The probability to populate an impurity from the source, 
and then to depopulate it towards the drain is directly related to the 
overlap of the atom-like impurity wave-functions with the corresponding
contacts leading to a conductance $G_1$ of this current channel
\begin{equation}
G_1 \propto \exp (-\frac{x}{a_B}) \exp (-\frac{d-x}{a_B})  
= \exp (-\frac{d}{a_B}).
\label{G1}
\end{equation}

This shows that in first approximation the conductance of this channel does not
depend on $x$. As shown above, a single-impurity channel already 
only selects impurities located within a very narrow range of $x$
close to the drain, Eq. (\ref{G1}) gives an additional argument for the
independence of this channel conductance $G_1$ on the actual location of
the impurity inside this thin layer near the drain.

\section{Two-, three-, multi-donor channel}

The above consideration shows, that due to the 
bend of the bottom of conduction
band following the transport voltage, there is no chance to notice current
flowing through a sequential chain of impurities (such as donors 1 and 3 in 
Fig. 2), connecting source and
drain. The contribution of such a chain will be totally masked by 
the current flowing directly
via the conduction band. The only way for multiple impurities to manifest
themselves in quantized conductance is to form multiple {\em parallel}
singe-impurity channels situated close enough to the drain as considered 
above.

Therefore, if $N>1$ impurities fall into the thin layer 
near the drain  to approximately the same $x$ coordinate 
as that of donor 1 in Fig. 2 (within the Bohr radius), we will see a
switching-on of an $N$-fold channel with conductance 
\begin{equation}
G_N=N G_1 \label{GN}
\end{equation}
at the same
threshold voltage $V_{eff}= V_1 = 0.525$ V as for a single-donor channels.

\section*{Discussion}

All the above considerations are only valid as long as the dot itself
can be regarded as an insulating system.
As the number of donors in a SQD grows, the dot becomes a metallic 
particle, and the 
conduction band edge in the dot aligns with the Fermi level of the 
electrodes.  
In a very simple estimate we define this transition to a metal when
the total volume of $N$ donors with an individual volume
of $4\pi/3 \times a_B^3$ exceeds the volume of the dot.  
This is an exaggerated version of the Mott criterion (\ref{Mott}) which holds
not only in bulk, but in a small structure, too.
For the analyzed example from above this gives
$N_{tot}=8$ as a limiting value. The   
practically interesting set $0,1,2,3,\ldots$ for both $N_{tot}$ and $N$
considered above is still far bellow this limit.

Quite a number of other mechanisms of electron transport might take place
in this system, the main one being resonant tunneling. Surprisingly, 
even taking into account such other mechanisms \cite{Kupriyanov} 
does not change much the main idea of the present paper.

\section*{Acknowledgments}

Useful discussions with M. Kupriyanov and S. Oda are gratefully acknowledged.
This work was supported in part by the Russian Foundation for Basic Research 
and by the Russian Program for Future Nanonelectronic Devices.

\newpage\bigskip
\noindent
{\bf Figure Captions}

\bigskip
\noindent
Fig. 1. Potential profile of the dot covered by an oxide layer at $V=0$. 
A donor is marked with a short bar.

\bigskip
\noindent
Fig. 2. Potential profile of the dot between effective source and drain 
biased with $V_{eff}=V_1$ (thick solid line) and $V_{eff}=V_{th}$
(dashed line).

\end{document}